\documentclass[12pt,preprint]{aastex}
\usepackage{color}
\usepackage{url}

\slugcomment{Accepted on 2015 December 29 for the publication in ApJ}

\shorttitle{Monitoring Observations of the Jupiter-Family Comet 17P/Holmes during 2014 Perihelion Passage}
\shortauthors{Yuna Kwon et al.}

\usepackage{amstext}

\begin{document}

\title{Monitoring Observations of the Jupiter-Family Comet 17P/Holmes during 2014 Perihelion Passage}

\author{Yuna Grace \textsc{Kwon}\altaffilmark{1}, Masateru \textsc{Ishiguro}}
\affil{Department of Physics and Astronomy, Seoul National University,\\
Gwanak, Seoul 151-742, Republic of Korea}
\altaffiltext{1}{ynkwon@astro.snu.ac.kr (Y.G.K)}

\author{Hidekazu \textsc{Hanayama}}
\affil{Ishigakijima Astronomical Observatory, National Astronomical
Observatory of Japan, \\ Ishigaki, Okinawa 907-0024, Japan}

\author{Daisuke \textsc{Kuroda}}
\affil{Okayama Astrophysical Observatory, National Astronomical
Observatory of Japan, \\ Asaguchi, Okayama 719-0232, Japan}

\author{Satoshi \textsc{Honda}, Jun \textsc{Takahashi}}
\affil{Nishi-Harima Astronomical Observatory, Center for Astronomy,
University of Hyogo, \\Sayo, Hyogo 679-5313, Japan}

\author{Yoonyoung \textsc{Kim}, Myung Gyoon \textsc{Lee}}
\affil{Department of Physics and Astronomy, Seoul National University,\\
Gwanak, Seoul 151-742, Republic of Korea}

\author{Young-Jun \textsc{Choi}, Myung-Jin \textsc{Kim}}
\affil{Korea Astronomy and Space Science Institute, 776 Daedeokdae-ro, \\
Yuseong-gu, 305-348 Daejeon, Republic of Korea}

\author{Jeremie J. \textsc{Vaubaillon}}
\affil{Observatoire de Paris, I.M.C.C.E., Denfert Rochereau, Bat. A.,
FR-75014 Paris, France}

\author{Takeshi \textsc{Miyaji}}
\affil{Ishigakijima Astronomical Observatory, National Astronomical
Observatory of Japan, \\ Ishigaki, Okinawa 907-0024, Japan}

\author{Kenshi \textsc{Yanagisawa}}
\affil{Okayama Astrophysical Observatory, National Astronomical
Observatory of Japan, \\Asaguchi, Okayama 719-0232, Japan}

\author{Michitoshi \textsc{Yoshida}}
\affil{Hiroshima Astrophysical Science Center, Hiroshima University,
\\Higashi-Hiroshima, Hiroshima 739-8526, Japan} 

\author{Kouji \textsc{Ohta}}
\affil{Department of Astronomy, Kyoto University, Sakyo, Kyoto 606-8502, Japan}

\author{Nobuyuki \textsc{Kawai}}
\affil{Department of Physics, Tokyo Institute of Technology
\\Meguro, Tokyo 152-8551, Japan}

\and
\author{Hideo \textsc{Fukushima}, Jun-ichi \textsc{Watanabe}}
\affil{National Astronomical Observatory, Mitaka, Tokyo
181-8588, Japan}

\begin{abstract}

We performed a monitoring observation of a Jupiter-Family comet, 17P/Holmes, during its 2014 perihelion passage to investigate its secular change in activity. The comet has drawn the attention of astronomers since its historic outburst in 2007, and this occasion was its first perihelion passage since then. We analyzed the obtained data using aperture photometry package and derived the $Af\rho$ parameter, a proxy for the dust production rate. We found that $Af\rho$ showed asymmetric properties with respect to the perihelion passage: it increased moderately from 100 cm at the heliocentric distance $r_\mathrm{h}$=2.6--3.1 AU to a maximal value of 185 cm at $r_\mathrm{h}$ = 2.2 AU (near the perihelion) during the inbound orbit, while dropping rapidly to 35 cm at  $r_\mathrm{h}$ = 3.2 AU during the outbound orbit.  We applied a model for characterizing dust production rates as a function of $r_\mathrm{h}$ and found that the fractional active area of the cometary nucleus had dropped from 20\%--40\% in 2008--2011 (around the aphelion) to 0.1\%--0.3\% in 2014--2015 (around the perihelion). This result suggests that a dust mantle would have developed rapidly in only one orbital revolution around the sun. Although a minor eruption was observed on UT 2015 January 26 at $r_\mathrm{h}$ = 3.0 AU, the areas excavated by the 2007 outburst would be covered with a layer of dust ($\lesssim$ 10 cm depth) which would be enough to insulate the subsurface ice and to keep the nucleus in a state of low activity.
\end{abstract}


\section{INTRODUCTION}
Comets are fossilized remnants of the formation epoch of the solar system, and most likely contain pristine ice and dust particles deep inside their bodies. On the other hand, their surfaces may have been altered by solar heating, galactic cosmic ray irradiation, and other processes \citep[see, e.g.,][]{Meech2004}. Cometary outbursts, which blow off the processed surface layers, can provide a unique opportunity to investigate how the fresh surface materials are evolved under the solar radiation field.  Among  known cometary outburst events, the 2007 outburst that occurred on the Jupiter-Family Comet 17P/Holmes was the most energetic outburst ever recorded \citep{Sekanina2009}.

17P/Holmes was discovered by Edwin Holmes on UT 1892 November 6 during an outburst \citep{Holmes1892}. Nearly one century later, it again underwent a similar, but more energetic, outburst on UT 2007 October 23.3, when the comet was on the outbound orbit at a heliocentric distance of $r_\mathrm{h}$ = $2.44$ AU. Its apparent flux increased by a factor of one million within about a day \citep{Sekanina2009,Hsieh et al.2010}. It is believed that a huge amount of dust grains (10$^{10}$--10$^{13}$ kg), corresponding to the mass of 1--100 m of the surface dust layer, was ejected from the comet \citep{Altenhoff et al.2009,Ishiguro et al.2010,Reach2010,Li et al.2011,Boissier et al.2012,Ishiguro et al.2013}. A signature of subsequent fragmentation was noticed through photometric analyses, indicating a continuous replenishment of cometary materials into the dust coma \citep{Stevenson et al.2010,Li et al.2011,Stevenson et al.2012}. The nucleus remained active for several years after the major outburst, showing minor outbursts on UT 2007 November 12 \citep{Stevenson et al.2012} and UT 2009 January 4.7 \citep{Miles2010}, and a persistent dust emission around its aphelion passage in 2010 \citep{Ishiguro et al.2013}. It was suggested that 17P/Holmes would make a spectacular appearance during the 2014 perihelion passage, because the fresh ice--dust conglomerate could be exposed by significant removal of the surface  layer \citep{Sekanina2009,Ishiguro et al.2013}.

Although vigorous research was conducted on the basis of observations just after the 2007 outburst, little is known about the long-term evolution in the activity of 17P/Holmes. The comet passed its perihelion in 2014 for the first time since the huge outburst, but so far, there have been no published reports about the physical state of the comet during the 2014 perihelion passage.  In this paper, we report our photometric monitoring observations of 17P/Holmes over the two years surrounding the 2014 perihelion passage using six ground-based telescopes at five observatories, motivated by the expectation by \citet{Ishiguro et al.2013} (hereafter paper I).  We describe the observations and data reduction in Section 2, and present the results in Section 3. We apply a simple model regarding the dust mass loss rate to understand the observed results, and discuss the evolution of the comet's nucleus under the solar radiation field in Section 4.

\section{OBSERVATIONS AND DATA REDUCTION} 
\subsection{Observations with Ground-Based Telescopes}
A summary of the observations is given in Table 1. In order to monitor the time variation in the activity as described above, we made observations of 17P/Holmes over about two years around the time of the perihelion passage on UT 2014 March 27 at five observatories, namely, the Ishigakijima Astronomical Observatory (hereafter, IAO), the Okayama Astrophysical Observatory (OAO), the Nishi-Harima Astronomical Observatory (NHAO), the Siding Spring Observatory (SSO), and the Bohyunsan Optical Astronomy Observatory (BOAO). In total, we obtained valid data at thirty-six epochs from UT 2013 May 4 to UT 2015 March 16. All telescopes were operated in a comet-tracking mode. Detailed information about the instruments and telescopes is shown below.

\subsubsection{IAO Observations}
A longterm monitoring observation was carried out mostly at IAO, which is located on Ishigaki island, Okinawa prefecture, Japan (24\arcdeg 22\arcmin 22.3\arcsec N, 124\arcdeg 08\arcmin 21.4\arcsec E, 197 m). Thanks to flexible operation as well as persistent effort by Dr. Hanayama (a co-author of the paper), we could obtain the greatest part of data (25 out of 36 nights) at IAO. We used the 105 cm Murikabushi telescope with the MITSuME instrument (g$'$, R$_\mathrm{C}$, and I$_\mathrm{C}$-band simultaneous imaging system, three 1,024 $\times$ 1,024 pixels CCDs with a 24.0 \micron\ pixel pitch). Each chip covers a field-of-view (FOV) of 12.3\arcmin $\times$ 12.3\arcmin\ with a pixel resolution of 0.72\arcsec. We obtained the data from UT 2014 July 2 through UT 2015 February 20.

\subsubsection{OAO Observations}
We made observation with the OAO 188 cm cassegrain telescope atop Mt. Chikurinji (34\arcdeg 34\arcmin 37.5\arcsec N, 133\arcdeg 35\arcmin 38.2\arcsec E, 372 m), in Okayama prefecture, Japan. The data were taken over two nights on UT 2014 September 16 and 20 as an open-use program (proposal ID: 14B-01). Moreover, we took snapshot images on UT 2015 March 16  during the engineering time of the observatory. We employed KOOLS (the Kyoto Okayama Optical Low-dispersion Spectrograph, 2,048$\times$4,096 pixels with a 15 \micron\ pixel scale) with a R$_\mathrm{C}$-band filter \citep{Yoshida2005}. In the $2 \times 2$ binning mode, the camera covers a $4.9\arcsec \times 4.9\arcsec$ FOV with a pixel resolution of 0.67\arcsec. In addition to these observations with KOOLS, we made use of the OAO 50 cm telescope at the same observatory on UT 2015 January 31. We employed MITSuME, which had an identical design to IAO/MITSuME. It has 26\arcmin$\times$26\arcmin\ pixels with a pixel resolution of 1.53\arcsec.

\subsubsection{NHAO Observation}
The NHAO, which is operated by Center for Astronomy, University of Hyogo,  is located atop Mt. Onade (35\arcdeg 01\arcmin 31.0\arcsec N, 134\arcdeg 20\arcmin08.0\arcsec E, 449 m), Sayo-cho, Hyogo prefecture, Japan. The observatory conducts scientific observations as well as scientific education and experience programs for the public. Our observations were carried out as a collaboration program with the observatory staff (Dr. Takahashi and Dr. Honda, co-authors of the paper). We made use of the Multiband Imager (MINT, 2,048$\times$2,064 pixels with a 15 \micron\ pixel scale) attached to the 200 cm Nayuta Telescope, at f/12 Cassegrain focal plane with R$_\mathrm{C}$-band filter. This combination covered  $10.9\arcmin \times 10.9\arcmin$ FOV with a pixel resolution of 0.32\arcsec. 

\subsubsection{SSO iTelescope Observations}
In addition, we made use of a remote commercial telescope, the iTelescope\footnote{http://www.itelescope.net/telescope-t30/} (which is available through the Internet) to monitor the magnitude of 17P/Holmes.  This observation was conducted as a part of the curriculum of an `Astronomical Observation' class for graduate students at Seoul National University. We made use of one of the multiple telescopes, T30, which consists of an optical CCD camera (FLI-PL6303E, 3,072$\times$2,048 pixels with a 9 \micron\ pixel scale) attached to a 50 cm telescope with a focal reducer. The telescope was located at and operated by the SSO (31\arcdeg 16\arcmin 24.0\arcsec S, 149\arcdeg 03\arcmin 52.0\arcsec E, 1165 m), New South Wales, Australia. The instrument has a FOV of $27.8\arcmin \times 41.6\arcmin$ with a pixel resolution of 0.81\arcsec\ without a binning mode. The observations at SSO were conducted for five nights in 2013 May--July when the telescopes at IAO, OAO, and NHAO were not available for our program.

\subsubsection{BOAO Observation}
Finally, we obtained data for one epoch on UT 2015 February 24 at BOAO. This observatory is located around the top of the Mt. Bohyun (36\arcdeg 9\arcmin 53.2\arcsec N, 128\arcdeg 58\arcmin 35.7\arcsec E, 1124 m), Yeongcheon-si, Gyeongsangbuk-do, Korea. We utilized the 180 cm cassegrain reflecting telescope with a 4k CCD camera (4,096 $\times$ 4,096 pixels), which covers $14.6\arcsec \times 14.6\arcsec$ and has a pixel resolution of 0.21\arcsec\ without a binning mode. This observation was conducted based on an unreviewed, web-based report of a minor eruption of the comet.
\subsection{Data Reduction}

We exploited only R$_\mathrm{C}$-band data because it is sensitive to the scattered light from cometary dust grains but is mostly uncontaminated by gaseous emissions at circum--nucleus. The observed raw data were preprocessed in a standard manner using dark (or bias if the CCD was cooled enough) and dome flat frames. To subtract background stellar objects and make composite images, we basically followed the method developed by \citet{Ishiguro2008}, tuning some parameters to erase the background sources neatly.  We calibrated the pixel coordinates into celestial coordinates using WCSTools \citep{Mink1997} or Astrometry.net software\footnote{http://astrometry.net}. In order to perform aperture photometry, we utilized the APPHOT package in IRAF. We conducted flux calibration using field stars listed in the UCAC-3 catalog \citep{Zacharias et al.2009}. We considered the calibration error of 0.1 mag for the UCAC-3 catalog, as well as photon noises from objects and sky background and readout noise in the CCDs. As a consequence, we could select thirty-six reliable epoch datasets with reasonable photometric errors ($<$0.15 mag).

\section{RESULTS}

\subsection{Appearances}
\begin{figure}
 \epsscale{1.0}
   \plotone{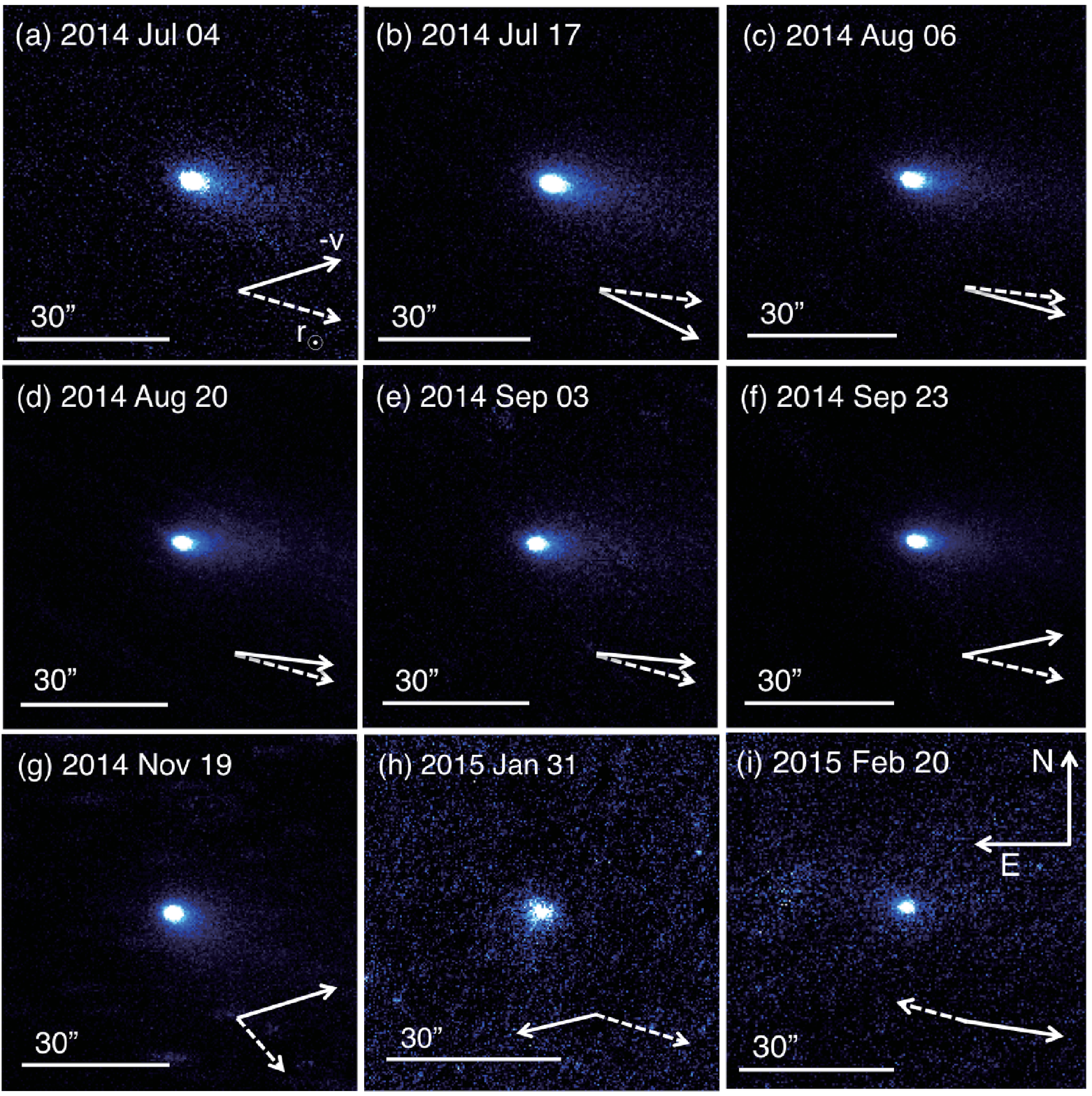}
  \caption{Time series of selected images of 17P/Holmes taken with an R$_\mathrm{C}$ filter. All images have the standard orientation, that is, north is up and east to the left. At the bottom right in each image, solid vectors denote the orientation of the negative velocity of the comet, and dashed vectors show radial vectors outward from the solar direction.}\label{fig:comping}     
\end{figure}

Figure \ref{fig:comping} shows selected R$_\mathrm{C}$-band images taken at IAO and OAO; the circumnuclear coma was clearly detected in the center of all images.  In addition, a feeble tail extended between the anti-solar direction ($\vec{r_\odot}$ vector) and the negative velocity vector projected on the celestial plane ($-v$). Note that the dust tail was not clearly seen in  Figure \ref{fig:comping} (h)--(i) because it extended along the line-of-sight (in fact, the vectors between the anti-solar direction and the negative velocity nearly aligned with the observer--comet vectors in these images). Unlike the images in Paper I and \citet{Stevenson et al.2014}, we could not detect a lingering dust tail oriented along $-v$. We interpret this invisibility as being due to the fact that the exposure times were not long enough to detect such a faint structure. In fact, our group succeeded in the detection of the lingering dust cloud by using much deeper imaging observation at the Kiso Observatory (Ishiguro et al. submitted to ApJ on 2015 September 12). We will report that result in a separate paper and  concentrate here on the analysis of fresh dust particles in the coma.

\subsection{Surface Brightness Profiles of the Dust Coma}
\begin{figure}
 \epsscale{0.9}
   \plotone{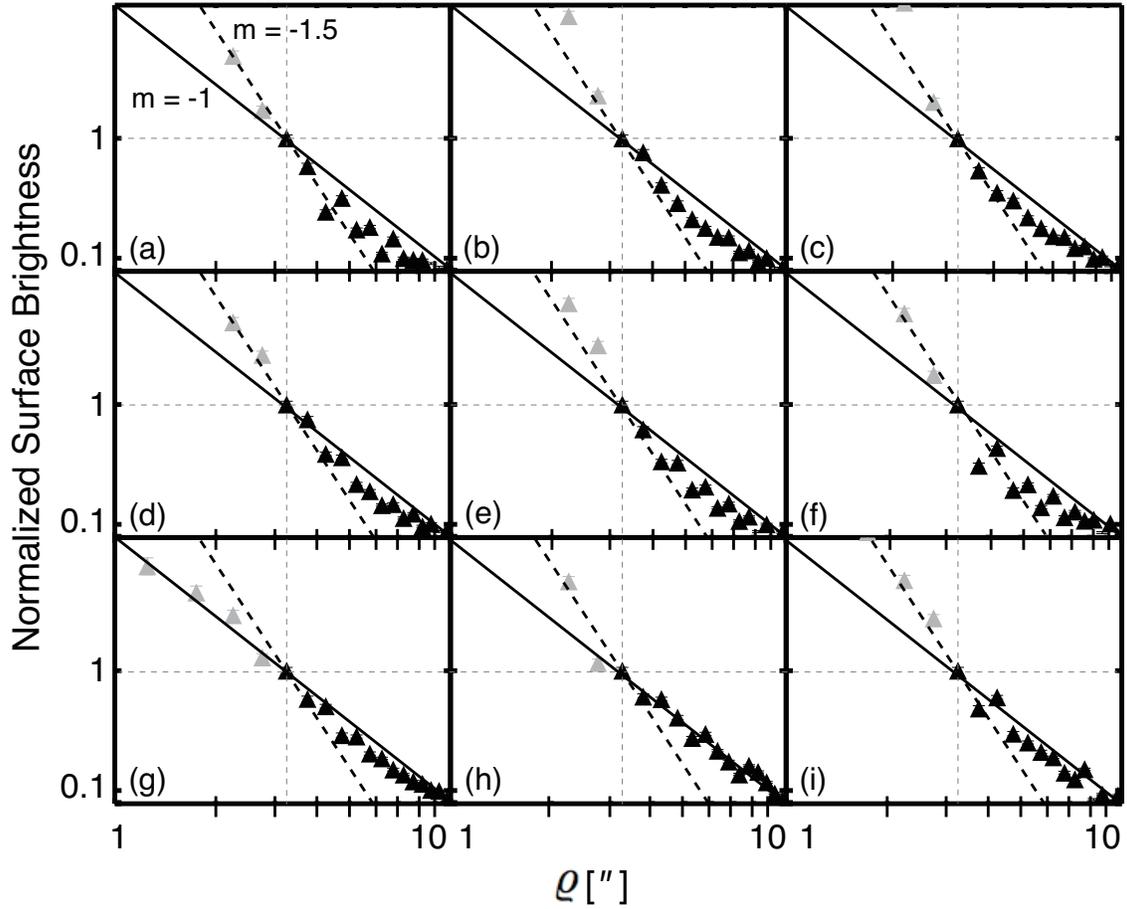}
   \caption{Normalized surface brightness profiles of 17P/Holmes as a function of the radial distance from the cometary nucleus. All brightnesses are normalized at $\varrho$ = 3.25$\arcsec$. From (a) to (i), each panel corresponds to the images listed in Figure 1. Upper dashed and lower solid lines exhibit gradients of -1.5 and -1, which show the ejection of dust particles from the nucleus without and with the radiation pressure of the Sun, respectively. For the fitting, we ignore the grey triangles whose radial distances are less than $\varrho$ = 3.25$\arcsec$.}\label{fig:radi}
    \end{figure}

Figure \ref{fig:radi} shows the surface brightness profiles of the near-nuclear coma with respect to the radial distance $\varrho$. We binned the radial-distance logarithmically spaced bins over 1.0\arcsec$<$$\varrho$$<$10.0\arcsec, calculated the average values within each bin, and plotted after normalization at $\varrho$ = 3.25\arcsec in which the data points are vulnerable to the seeing effects. For comparison, we show power functions with the indices of $m$ = -1 (solid lines) and $m$ = -1.5 (dashed ones). It is found that our data show profiles lying between these two power functions, but strictly speaking, closer to the power function with $m$=-1.5 (except for the later stages at which the comet was located nearly 3 AU from the Sun).

It is known that a steady-state dust cloud expanding with an initial velocity but no acceleration shows a surface brightness profile with $m$ = -1. The application of the $A f \rho$ parameter (where $A$ denotes the albedo of dust particles and $f$ their packing density within a considered aperture, $\rho$) was contrived assuming a simple steady-state dust coma. In this case, the flux within the aperture, multiplied by $\rho$, should be a constant regardless of the size \citep{A'Hearn et al.1984}. However, this is not the case for general cometary dust comae, where dust particles are supposed to be accelerated by the solar radiation pressure. \citet{Jewitt1987} considered a more realistic case of a cometary dust cloud, that is, a steady-state flow of dust particles under the solar radiation field, and noticed that the surface brightness follows a power function with an index of $m$ = -1.5. Thus, the surface brightness profiles of 17P/Holmes can be explained by continuous dust ejection from the nucleus under the solar radiation field. Since the solar radiation pressure made a non-negligible contribution to the surface brightness profile of 17P/Holmes, we chose a small physical distance, $\rho$ = 5,000 km (4.28\arcsec--7.55\arcsec\ depending on the distance from the comet), for the aperture photometry below. This value is large enough to diminish the nightly atmospheric effect while small enough to reduce the influence of solar radiation pressure.

\subsection{Coma Absolute Magnitude and $Af\rho$ }

\begin{figure}
 \epsscale{0.8}
   \plotone{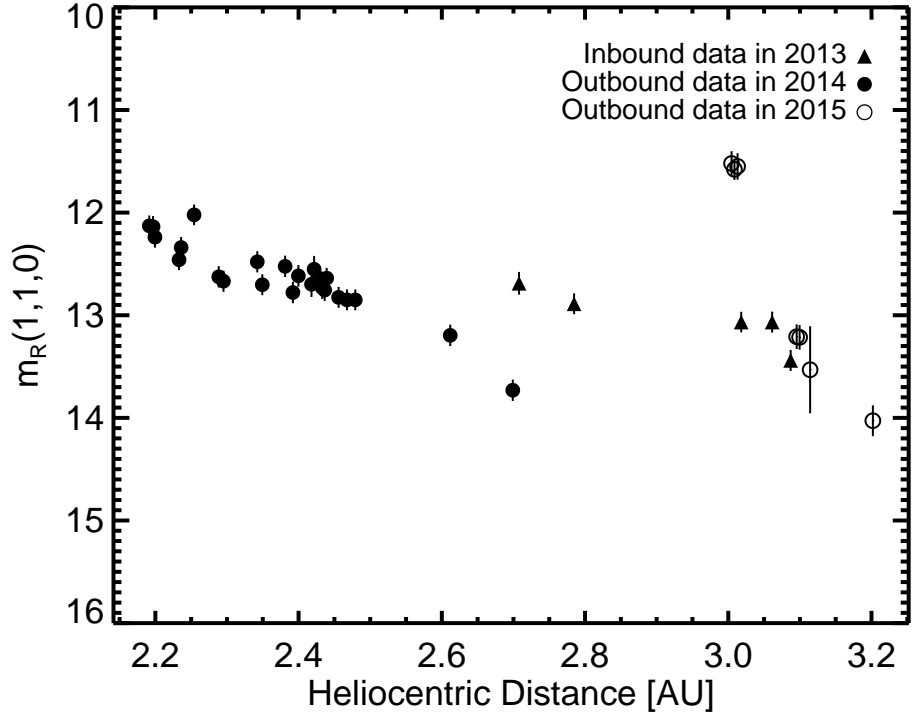}
  \caption{Absolute magnitudes of 17P/Holmes  ($m_\mathrm{R} (1,0,0)$) as a function of the heliocentric distances  ($r_\mathrm{h}$). Triangle and filled-circle labels denote the data obtained during the 2013 inbound and 2014 outbound orbits, respectively; open circles denote the sudden brightness enhancement that occurred on UT 2015 January 26. For all occasions, we extracted the magnitude within the aperture size corresponding to a 5,000 km radial distance from the cometary center. }
 \label{fig:OPTimage} 
\end{figure}

We deduced the absolute magnitudes of the comet, which corresponds to the magnitude at a hypothetical point in space, namely, a heliocentric distance of $r_\mathrm{h}$ = 1 AU,
an observer distance of $\Delta$ = 1 AU, and a solar phase (sun--comet--observer's) angle of $\alpha$ = 0\arcdeg. This magnitude is given by,
\begin{eqnarray}
m_\mathrm{R}(1,1,0)=m_\mathrm{R} - 5~\log_{10}(r_\mathrm{h} \Delta) - 2.5\log_{10} \Phi(\alpha),
\label{eq:eq1}
\end{eqnarray}
\noindent where $m_\mathrm{R}$ denotes the observed R$_\mathrm{C}$-band magnitude. Since the comet was observed in a moderate phase angle range ($\alpha$ = 12--25\arcdeg), we need to correct the phase darkening of the dust cloud as written in the third term of Eq. (\ref{eq:eq1}).  We adopted  a commonly-used empirical scattering phase function, 2.5$\log_{10}$$\Phi$($\alpha$) = $\beta$$\alpha$, where $\beta=0.035$ mag deg$^{-1}$ is assumed \citep[see e.g.,][]{Lamy et al.2004}. Figure \ref{fig:OPTimage} shows the absolute magnitudes after UT 2008 December 24 with respect to $r_\mathrm{h}$. For comparison, we appended the absolute magnitudes beyond 4 AU from Paper I.
In Figure \ref{fig:OPTimage}, there is a general trend of the magnitude decreasing (i.e. the brightness increasing) as the comet approaches the perihelion. The trend can be explained naturally by extra-solar heating making the comet more active.  A temporal increase in the brightness (i.e. a minor outburst) was detected on UT 2015 January 26 at r$_\mathrm{h}$ = 2.99 AU. Except for the data during this minor outburst, the outbound data (in 2014--2015) appear to be about 1--2 magnitude fainter than the inbound data (in 2013), although they were taken around the same distance from the Sun. 

We derived $Af\rho$ as well, which is given by,

\begin{displaymath}
Af\rho = \mathrm{K}~\biggl[\frac{\Delta}{\mathrm{AU}}\biggr]^2\biggl[\frac{r_\mathrm{h}}{\mathrm{AU}}\biggr]^2 \biggl[\frac{\rho}{\mathrm{cm}}\biggr]^{-1} \times {2.512^{(-m_\odot + m_\mathrm{R})}},
\label{eq:eq2}
\end{displaymath}

\noindent where $m_\odot$ is the R$_\mathrm{C}$-band magnitude of the sun at $r_\mathrm{h}$ = 1 AU.  $\mathrm{K}$ is a constant of 8.95$\times$$10^{26}$, used for unitary transformation of distances. We adopted $m_\odot$$ = -27.11$ \citep{Drilling2000}. We tabulated the derived $Af\rho$ values and absolute magnitudes in Table 2.

\begin{figure}
 \epsscale{0.8}
   \plotone{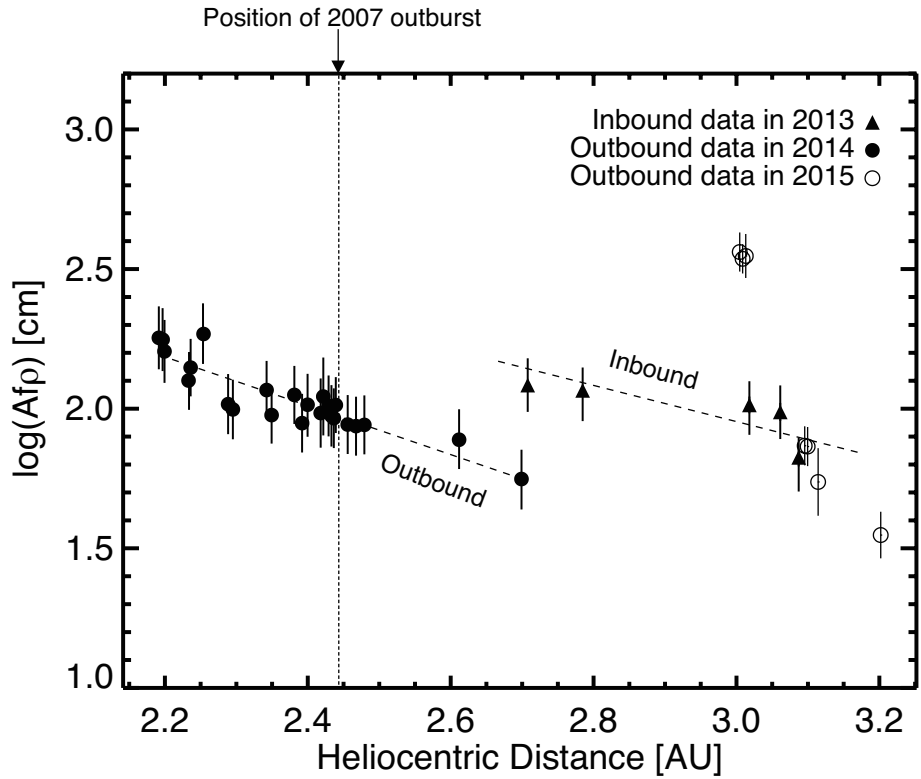}
  \caption{$Af\rho$ values taken with a logarithmic scale with respect to the heliocentric distances  ($r_\mathrm{h}$). The vertical short-dashed line denotes the exact position of 2007 outburst and the dashed lines show the logarithmic plots that follow the decreasing log($Af\rho$) trend of the data.}\label{fig:Afrho} 
\end{figure}

Figure \ref{fig:Afrho} shows the $r_\mathrm{h}$-dependence of $Af\rho$ between 2013 and 2015. In this figure, the notations of all symbols are the same as those in Figure \ref{fig:OPTimage}. At the given heliocentric distance, $Af\rho$ values in the 2014 data (after the perihelion passage) are at least 2.5 times below those in 2013 (before the perihelion passage). We appended the trend lines with a slope in the single logarithmic plot ($a$) to signal the differences in $Af\rho$ between the  inbound and outbound orbits. By fitting data with a function, $A f \rho \propto 10^{a r_\mathrm{h}}$, we obtained $a$ = -0.16 cm AU$^{-1}$ for inbound data and -0.20 cm AU$^{-1}$ for outbound data. We did not detect any change in $A f \rho$ at the exact position of the 2007 outburst (see the dashed vertical line). On the contrary, a sudden but minor brightening was observed on UT 2015 January 26 (at $r_\mathrm{h}$ = 2.99 AU ), increasing the flux by a factor of $\sim$10. Soon after the minor outburst, $Af\rho$ was higher than the trend line for about one month and came back again to follow the original decreasing trend, although the trend lines do not have any physical implication.

\section{DISCUSSION}

In this section, we provide an interpretation of the secular change in the comet activity (described in Section 3) by computing the dust mass loss rate and fraction of the active area of the cometary nucleus, and substantiate the fact that 17P/Holmes shows a rapid development of a dust mantle on its nuclear surface.

\subsection{Estimation of the Dust Mass Loss Rate }

As suggested in \citet{A'Hearn et al.1984}, $Af\rho$ can be used as an index to characterize the dust production rate. It is also possible to derive the dust production rate from photometric magnitudes. We first calculated the mean optical cross section of the dust coma by adapting the same manners as \citet{Luu1992} whose equations are defined as
\begin{eqnarray}
p_\mathrm{R}~C_\mathrm{c} = 2.25\times10^{22}\pi~r_\mathrm{h}^2~\Delta^2~10^{0.4({m_\odot} - \bar{m}_\mathrm{R})},
\label{eq:eq3}
\end{eqnarray}
\noindent where $p_\mathrm{R}$ is the geometric albedo ($p_\mathrm{R}$=0.04 was assumed) for 17P/Holmes, and  $C_\mathrm{c}$ is the optical cross-section of dust particles within the aperture. Next, we obtained a parameter $\eta$, defined as the ratio of $C_\mathrm{c}$ to the cross section of a nucleus, $C_\mathrm{n}$ (i.e. $\eta$ = $C_\mathrm{c}$/$C_\mathrm{n}$). Applying an effective nuclear radius of $r_\mathrm{obj}$ = 2.07$\pm$0.31 km derived by an infrared observation \citep{Stevenson et al.2014}, we obtained $C_\mathrm{n}$  = 1.34 $\times$ 10$^{7}$ m$^2$. Except for the sudden brightening at $r_\mathrm{h}$ = 2.99 AU, $\eta$ has decreased from 5.38 to 1.11, which is even smaller than the value measured at $r_\mathrm{h}$ = 4.15 AU on UT 2008 December 26 in Paper I.

We derived the dust mass loss rate with $\eta$ in the same manner as Paper I, using an equation in \citet{Luu1992},
\begin{eqnarray}
\dot{M_\mathrm{d}} = \frac{1.1 \times 10^{-3} \pi \rho_\mathrm{d} \bar{a} \eta r_\mathrm{obj}^{2}}{\phi r_\mathrm{h}^{1/2} \Delta}.
\label{eq:eq4}
\end{eqnarray}
\noindent where $\rho_\mathrm{d}$ is the mass density of dust particles, $\bar{a}$ for the averaged small particle size, $\phi$ for the aperture size for photometry in arc seconds, respectively. In estimating $\dot{M_\mathrm{d}}$, we assumed  $\bar{a}$ = 1.0 $\times$ $10^{-6}$ m and $\rho_\mathrm{d}$ = $1,000$ kg m$^{-3}$. The effects of different aperture sizes can be cancelled by the denominator in Eq.(\ref{eq:eq4}). 

\begin{figure}
 \epsscale{0.8}
   \plotone{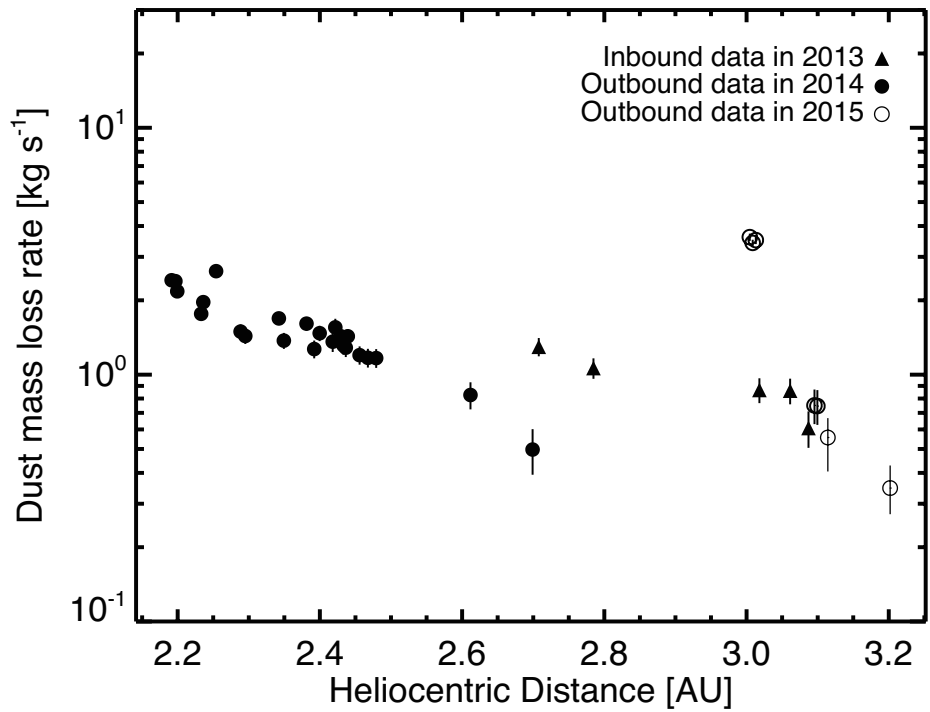}
  \caption{The dust mass loss rate ($\dot{M_\mathrm{d}}$) of 17P/Holmes during our observations with respect to the heliocentric distances ($r_\mathrm{h}$). Both inbound and outbound data show a rapid decrease of $\dot{M_\mathrm{d}}$, clearly distinguished from the results of Paper I around the aphelion, and $\dot{M_\mathrm{d}}$ in 2014--2015 are a factor of two lower than those in 2013.}\label{fig:Md} 
\end{figure}

Figure \ref{fig:Md} shows the result for the dust-production rate. The change of the dust mass loss rate appears to be explained in a similar way to those of the absolute magnitude and of the $Af\rho$, in that it has been declining with increasing heliocentric distance. The dust production rates around the 2014 perihelion passage were about five orders of magnitudes lower than the maximum value during the 2007 outburst \citep[3$\times$10$^5$ kg s$^{-1}$,][]{Li et al.2011}, while they were  comparable to that of the pre-outburst data at $r_\mathrm{h}$ = 2.23 AU (i.e. 2.8 kg s$^{-1}$ in paper I). The peak around $r_\mathrm{h}$ = 3.0 AU is in accordance with the sudden explosion of 17P/Holmes.

It is true that the derivation of the above dust mass loss rate is crude because we have not considered the dust average particle size ($\bar{a}$) seriously. In fact, the ejection of large particles has been confirmed from a variety of comets, including 17P/Holmes (see paper I).  Our estimate for $\dot{M_\mathrm{d}}$ would be significantly (2--3 orders of magnitude) lower than in the case of big particles ($\bar{a}$ = 100 \micron--1 mm). However,  we suggest that the mass loss derivation still yields valuable information for monitoring the activity of the comet. It would not be unreasonable to assume that the size distribution would be constant over the period in a steady state. In this case, the relative values of $\dot{M_\mathrm{d}}$ are reasonable for comparing the activity at different solar distances.

It is interesting to notice that the dust production rate became equivalent to the level before the 2007 outburst when the {\it AKARI} infrared telescope happened to observe the comet (see paper I). Note that almost the same model was applied to derive $\dot{M_\mathrm{d}}$ using the {\it AKARI} infrared data, although there is a subtle difference in conversion from flux to the cross-section between optical and infrared data. This similarity suggests that the activity was restored to its former state, although the comet has shown lingering activity for years when it was around its aphelion. It is probable that the area excavated by the 2007 outburst was mostly covered with an insulating dust mantle.
\\
\subsection{Fractional Active Area of the Nucleus}

Now, we show our derivation of the active area fraction of the nuclear surface. The overall methodology is the same as that of Paper I (presented in Section 3.4 therein). We consider an instant thermal equilibrium between the incident solar influx,  thermal radiation from the nucleus, and latent heat of water ice sublimation to calculate the gas production rate at given heliocentric distances. The energy balance equation of a spherical body is given by
\begin{eqnarray}
\frac{S_\mathrm{0}}{r_\mathrm{h}^2}(1-A_\mathrm{p}) \cos ~z=\epsilon_\mathrm{E} \sigma T^4 + L_\mathrm{w}(T)\frac{dZ}{dt}~~
 \label{eq:eq5}
\end{eqnarray}
\noindent where $S_0$ is the solar flux at $r_\mathrm{h}$ = 1AU, $z$ is the zenith distance of the sunlight (the angle between the sunlight and normal vector of the local surface), $\epsilon_\mathrm{E}$ is the emissivity, $\sigma$ is the Stefan-Boltzmann constant, $T$ is the equilibrium temperature of the nucleus, $L_\mathrm{w}(T)$ is the latent heat of water ice sublimation, and ${dZ}/{dt}$ is the sublimation rate of water ice. 
$A_\mathrm{p}$ is the $R_\mathrm{C}$-band geometric albedo identical to $p_\mathrm{R}$. Since $L_\mathrm{w}(T)$ and ${dZ}/{dt}$ (as well as the sticking coefficient and the Clausius--Clapeyron equation) are dependent on the equilibrium temperature ($T$), we can rearrange Eq. (\ref{eq:eq5}) into a function of equilibrium temperature. Integrating ${dZ}/{dt}$ over the sunlit hemisphere with the calculated temperature and assuming a water/gas ratio of $\kappa$, we can compute the dust mass loss rate:
\begin{eqnarray}
\dot{M}_\mathrm{d}=\frac{2 \pi r_\mathrm{obj}^2 f }{\kappa}\int_0^{\pi/2} \left(\frac{dZ}{dt}\right) \sin z \,dz ,
 \label{eq:eq6}
\end{eqnarray}
\noindent
where $f$ is the active area fraction on the cometary surface. We adopted $\kappa$ = 0.6, which provided reasonable results from observations (paper I) and a theoretical thermal model of the 17P/Holmes nucleus \citep{Hillman2012}.

\begin{figure}
 \epsscale{1.0}
   \plotone{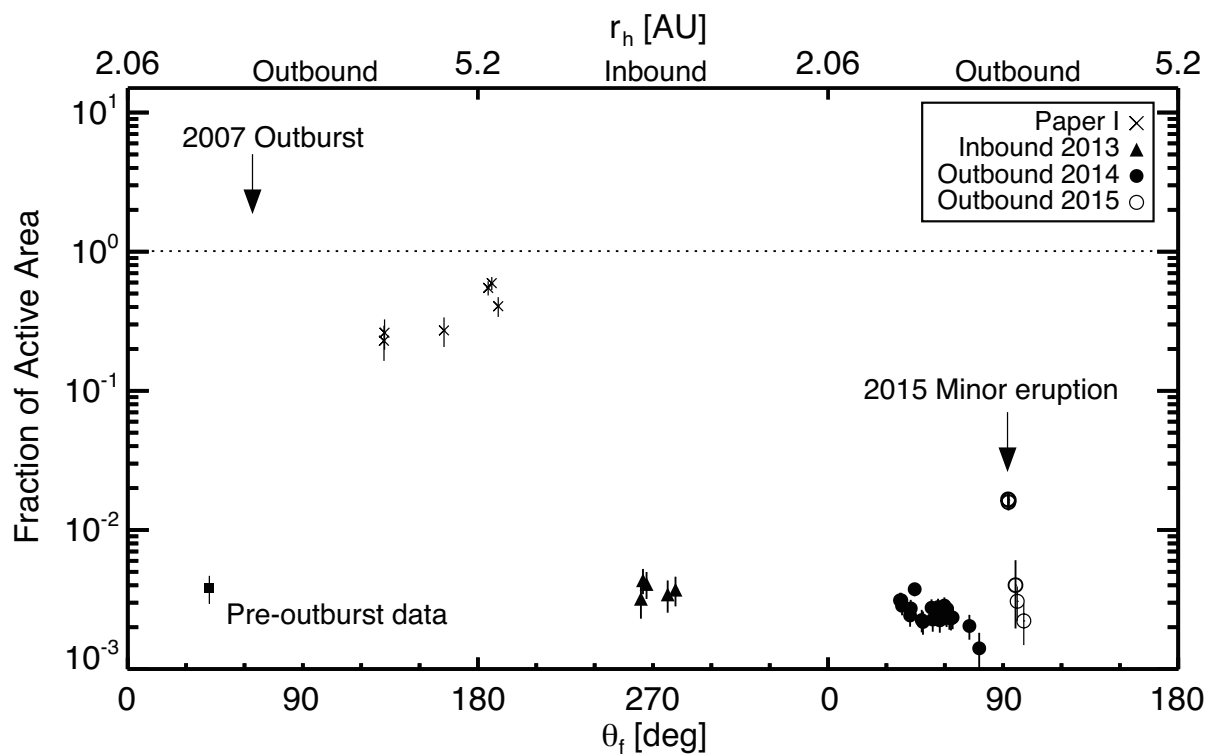}
  \caption{Secular evolution of the fractional active area ($f$) of the cometary nucleus as a function of the true anomaly ($\theta_\mathrm{r}$). The upper-left arrow denotes the major outburst of 17P/Holmes in 2007. The dashed lines present the average value of $f$ during the periods of the 2008--2011 outbound, 2013 inbound, and 2014 outbound orbits. We also quote the pre-outburst data (filled-square) to compare the activity rate of the comet.}\label{fig:f} 
\end{figure}

Comparing the derived dust mass loss rates  in Eq. (\ref{eq:eq4}) with theoretical ones in Eq. (\ref{eq:eq6}), we could derive the fractional active area of the cometary nucleus, $f$. In Figure \ref{fig:f}, $f$ is plotted with respect to the true anomaly, $\theta_\mathrm{f}$. Since the true anomaly of a comet changes slowly around the aphelion but rapidly near the perihelion, this figure efficiently shows the time evolution of the fractional active area. Crosses and filled squares are quoted from Paper I to compensate for the fractional active area soon before and after the 2007 outburst event. After the 2007 outburst, the ratio of active-to-inert surface area had significantly increased, even around the aphelion, probably because the excavated area had been sufficiently preserved under low temperature to quench the water sublimation rate. During the inbound orbit, however, the $f$ value significantly decreased by nearly two orders of magnitude and became equivalent to the level of activity before the 2007 outburst. Although the fractional active areas were tentatively induced by the sudden brightness enhancement of UT January 26 at 2.99 AU, it fell to the original level within a month. From this evidence, we conjecture that adiathermic materials (sublimation-driven dust particles) spread over the surface and prevented the sublimating icy volatile materials from being ejected outward from the nucleus.

\subsection{Rapid Development of the Dust Mantle}

Up to here, we have shown a rapid dormancy for 17P/Holmes, due to the formation of an insulating dust mantle, despite an expectation in Paper I. To sum up the major finding of this research, we consider whether the period is enough to develop such a dust mantle on the comet.

\citet{Rickman et al.1990} implied that the formation of a stable dust mantle is preferred when the spin axis is on the comet's orbital plane and when the perihelion distance is $r_h$ $\ge$ 2 AU. \citet{Jewitt2015} suggested that solar heat penetrates only a few diurnal thermal skin depths into the surface, about 10--20 cm on 67P/Churyumov--Gerasimenko (Jupiter Family Comet, a dynamical analog to 17P/Holmes).  In those regards, it is noteworthy that 17P/Holmes has shown saliently dormant features over intervals far shorter than the dynamical lifetime \citep[$\approx$ 1$\times$10$^{4}$ years,][]{Levison1997}.
Since a large fraction of the fresh surface of 17P/Holmes was exposed by the 2007 outburst, this would provide a unique opportunity for researchers to estimate the formation timescale of the dust mantle. 17P/Holmes spends more than half a revolution period beyond the heliocentric distance of 3 AU. Taking into account the persistent activities of the comet around its aphelion (Paper I), it should have had favorable conditions in developing the dust mantle with continuously outward sublimating volatiles over quite large portions of its orbit. 

We performed an order-of-magnitude estimate following a method by \citet{Jewitt2002} (Subsection 4.7 therein). Regarding the ballistic redeposition of resurfacing of a comet and restricting the activity to within a limited active area, we considered a rotational period of $P_{rot}$ = 7.2-- 12.8 h \citep{Snodgrass et al.2006}, a nuclear density of $\rho_{n}$ = 1,000 kg m$^{-3}$, a capture fraction of $f_{B}$ = $10^{-3}$, an active area fraction of the comet's surface ($\psi$ in \citet{Jewitt2002}) of $f$ =  (2--3) $\times~10^{-3}$, and a mass loss rate of $\dot{M}_\mathrm{d}$ (see Subsection 4.2). As a result, we obtained a diurnal skin depth of the dust mantle of $L_D$ = 5--7 cm, thick enough to quench subsurface ice sublimation, and found that the timescale for growing up to $L_D$ was $\tau_{B}$ = 5--10 years. Although there may be uncertainties in this estimate, our result (i.e. a mantle growth timescale of $\lesssim$7 years) is consistent with the estimate by the ballistic mantle model.

There are several examples of {\it in situ} observations on similar comets that support the rapid formation of a dust mantle on Jupiter--Family comets. The nucleus of 9P/Tempel 1 has been investigated by two flyby spacecraft missions, Stardust--NExT and Deep Impact, separated by one orbital period of 5.5 years \citep{Kobayashi et al.2013,Thomas et al.2013}. From the topographical changes that took place between the missions and observations of the jet plume, it was suggested that 9P/Tempel 1 has a dust mantle consisting of micron-scale dust grains. In addition, \citet{Schulz et al.2015} investigated data for 67P/Churyumov--Gerasimenko, taken with the Cometary Secondary Ion Mass Analyser (COSIMA) onboard the ESA Rosetta spacecraft, and suggested that ice-free dust would have accumulated on the cometary surface, quenching the subsurface icy volatile sublimations and restricting the activity within the localized active area.

\citet{Rosenberg2009} studied a numerical model of 67P/Churyumov--Gerasimenko and found that the dust mantle thickness varies over the surface in the range of 1--10cm, which is slightly thinner and thicker than the diurnal skin depth of solar radiation at a given heliocentric distance. In addition, a state of low activity can exist ubiquitously over cometary surfaces, mainly due to thermal effects that help active regions to keep opening up again \citep{Rickman et al.1990}. In these regards, if the thickness of 17P/Holmes's surface coverage was shallow enough for sunlights to penetrate the refractory layers or for inner sublimation pressure to crack and come to the dust--ice interface, certain regions could suddenly manifest minor outbursts. Therefore, minor brightness enhancement of 17P/Holmes on UT 2015 Jan 26 can be explained in terms of an unstable nucleus and inhomogeneous dust mantle coverage.

We also note that the post-outburst activity of 17P/Holmes  in 1892 showed a trend similar to our results. The flux dropped rapidly by two orders magnitude in the following return in 1899 but slowly decreased by a factor of 2--5 over one century \citep{Sekanina2009}, suggesting that the rapid drop could be explained by the rapid development of dust mantle as we discussed above, while the prolonged hibernation of the cometary activity can be attributed to the additional accumulation of refractory materials and/or lost of near-surface ice. From the repetitive evolutionary pattern of 17P/Holmes over one century, it is reasonable to think that the activity of the comet is highly controlled by the formation and evolution of dust mantle.

\section{SUMMARY}
In this paper, we performed monitoring observations of the comet 17P/Holmes to examine the activity rate of the comet before and after its perihelion passage on UT 2014 March 27.57 over a period of nearly two years from May 2013 to March 2015. Our work leads us to the following principal conclusions:

\begin{enumerate}

\item  $Af\rho$ showed asymmetric properties with respect to the perihelion passage. It increased moderately from 100 cm at a heliocentric distance of $r_\mathrm{h}$ = 2.6--3.1 AU to a maximal value of 185 cm at $r_\mathrm{h}$ = 2.2 AU (near the perihelion) during the inbound orbit, while dropping rapidly to 35 cm at  $r_\mathrm{h}$ = 3.2 AU during the outbound orbit.

\item  The dust mass loss rate of the inner dust coma of 17P/Holmes has been declining with increasing heliocentric distance, similar to the above-mentioned physical quantities. Compared to the values taken right after the 2007 outburst \citep{Li et al.2011,Stevenson et al.2012}, our results present a dust production rate that has been utterly quenched by about five orders of magnitudes.

\item  Secular evolution of the fractional active area over the cometary surface as a function of true anomaly shows that the overall activity of 17P/Holmes has been significantly restrained by nearly two orders of magnitude, representing a decrease  from 20\% -- 40\% (Paper I) to 0.1\% -- 0.3\% in the 2014--2015 outbound orbits.

\item As an order-of-magnitude estimation, we calculated the diurnal skin depth and growth timescale of the dust mantle of 17P/Holmes. The dust mantle, which is now nearly 5--7 cm thick, has been developing apace over the $\sim$ 7 years.

\end{enumerate}

Our observations have proved that refractory dust mantles effectively suppress cometary activities, which are induced by sublimating outward icy volatiles, within a very short time interval. Furthermore, this result might be indicative of the fact that defunct (i.e. inactive and/or dormant) comets are aided by the development of their dust mantles acquired over their evolutionary histories in the inner part of the Solar System.

\acknowledgments
\noindent
This work at Seoul National University was supported by a National Research Foundation of Korea (NRF) grant funded by the Korean Government (MEST) (No. 2012R1A4A1028713). The observations at OAO, IAO, and NHAO were partially supported by the Ministry of Education, Culture, Sports, Science and Technology of Japan, Grant-in-Aid Nos. 23340048, 24000004, 24244014, and 24840031, 14GS0211, and 19047003. Yuna Kwon was supported by the Global PH.D Fellowship Program through the National Research Foundation of Korea (NRF) funded by the Ministry of Education (NRF-2015H1A2A1034260). In addition, Y.J.C. and M.J.K. were supported by the National Research Council of Science \& Technology.


\clearpage


\clearpage


\clearpage


\clearpage

\clearpage


\clearpage


\clearpage


\clearpage
\begin{deluxetable}{lrrrrrrrrr}
\tablecaption{Journal of 17P/Holmes Monitoring Observations}
\label{tab:observation}
\tabletypesize{\footnotesize}
\tablehead{\colhead{UT date} & \colhead{Telescope (Instrument)} & \colhead{Filter} & \colhead{Exptime} & \colhead{$N$} & \colhead{$\phi$ [\arcsec] } & \colhead{$r_\mathrm{h}$} & \colhead{$\Delta$} & \colhead{$\alpha$} & \colhead{$\theta_\mathrm{T}$} }
\startdata
 \\
2010/10/14.25 &  Aphelion & \ldots & \ldots  & \ldots & \ldots & 5.19 & \ldots & \ldots  & 180.0\\
 \\
2013/05/04.67 & SSO050 (FLI-PL6303E)   & R$_\mathrm{C}$ & 120 & 10 & 5.86 & 3.09 & 2.35 & 14.7 & 263.8\\
2013/05/10.75 & SSO050 (FLI-PL6303E)   & R$_\mathrm{C}$ & 120 & 5   & 6.06 & 3.06 & 2.27 & 13.8  & 264.9\\
2013/05/20.54 & SSO050 (FLI-PL6303E)   & R$_\mathrm{C}$ & 120 & 10 & 6.39 & 3.02 & 2.16 & 12.0  & 266.7\\
2013/07/13.58 & SSO050 (FLI-PL6303E)   & R$_\mathrm{C}$ & 120 & 6   & 7.17 & 2.78 & 1.92 & 13.4  & 277.6\\
2013/07/31.58 & SSO050 (FLI-PL6303E)   & R$_\mathrm{C}$ & 100 & 5   & 6.94 & 2.71 & 1.99 & 17.7  & 281.6\\
\\
2014/03/27.56     &  Perihelion  & \ldots &      \ldots        & \ldots & \ldots & 2.05 & \ldots & \ldots  & 0.0\\
\\
2014/07/02.79 & IAO105 (MITSuME) & R$_\mathrm{C}$ & 180 & 7   & 5.19 & 2.19 & 2.66 & 21.6 & 37.3\\
2014/07/04.79 & IAO105 (MITSuME) & R$_\mathrm{C}$ & 180 & 7   & 5.21 & 2.20 & 2.64 & 21.8 & 37.8\\
2014/07/05.75 & IAO105 (MITSuME) & R$_\mathrm{C}$ & 180 & 12 & 5.23 & 2.20 & 2.64 & 22.0 & 38.0\\
2014/07/17.75 & IAO105 (MITSuME) & R$_\mathrm{C}$ & 180 & 12 & 5.37 & 2.23 & 2.57 & 23.1 & 42.2\\
2014/07/18.76 & IAO105 (MITSuME) & R$_\mathrm{C}$ & 180 & 14 & 5.38 & 2.24 & 2.56 & 23.2 & 42.5\\
2014/07/24.81 & IAO105 (MITSuME) & R$_\mathrm{C}$ & 180 & 13 & 5.46 & 2.25 & 2.52 & 23.7 & 44.5\\
2014/08/04.76 & IAO105 (MITSuME) & R$_\mathrm{C}$ & 180 & 18 & 5.63 & 2.29 & 2.45 & 24.4 & 48.1\\
2014/08/06.79 & IAO105 (MITSuME) & R$_\mathrm{C}$ & 180 & 12 & 5.66 & 2.30 & 2.44 & 24.5 & 48.8\\
2014/08/20.80 & IAO105 (MITSuME) & R$_\mathrm{C}$ & 180 & 22 & 5.90 & 2.34 & 2.34 & 25.0 & 53.2\\
2014/08/22.79 & IAO105 (MITSuME) & R$_\mathrm{C}$ & 180 & 2   & 5.93 & 2.35 & 2.32 & 25.0 & 53.8\\
2014/08/31.72 & IAO105 (MITSuME) & R$_\mathrm{C}$ & 180 & 18 & 6.11 & 2.38 & 2.26 & 25.0 & 56.5\\
2014/09/03.71 & IAO105 (MITSuME) & R$_\mathrm{C}$ & 180 & 9   & 6.17 & 2.39 & 2.24 & 24.9 & 57.4\\
2014/09/05.80 & IAO105 (MITSuME) & R$_\mathrm{C}$ & 180 & 19 & 6.21 & 2.40 & 2.22 & 24.8 & 58.0\\
2014/09/10.73 & IAO105 (MITSuME) & R$_\mathrm{C}$ & 180 & 12 & 6.32 & 2.42 & 2.18 & 24.6 & 59.5\\
2014/09/11.70 & IAO105 (MITSuME) & R$_\mathrm{C}$ & 180 & 26  & 6.34 & 2.42 & 2.18 & 24.6 & 59.8\\
2014/09/13.67 & IAO105 (MITSuME) & R$_\mathrm{C}$ & 180 & 16  & 6.38 & 2.43 & 2.16 & 24.4 & 60.4\\
2014/09/14.74 & IAO105 (MITSuME) & R$_\mathrm{C}$ & 180 & 14  & 6.41 & 2.43 & 2.15 & 24.4 & 60.6\\
2014/09/15.70 & IAO105 (MITSuME) & R$_\mathrm{C}$ & 180 & 15  & 6.43 & 2.44 & 2.15 & 24.3 & 60.9\\
2014/09/16.33 & OAO188 (KOOLS)  & R$_\mathrm{C}$ & 120 & 12   & 6.44 & 2.44 & 2.14 & 24.3 & 61.3\\
2014/09/20.75 & OAO188 (KOOLS)  & R$_\mathrm{C}$ & 120 & 71   & 6.54 & 2.46 & 2.11 & 23.9 & 62.3\\
2014/09/23.76 & IAO105 (MITSuME) & R$_\mathrm{C}$ & 180 & 40  & 6.61 & 2.47 & 2.08 & 23.6 & 63.2\\
2014/09/26.75 & IAO105 (MITSuME) & R$_\mathrm{C}$ & 180 & 25  & 6.68 & 2.48 & 2.06 & 23.3 & 64.0\\
2014/10/29.68 & IAO105 (MITSuME) & R$_\mathrm{C}$ & 180 & 16  & 7.37 & 2.61 & 1.87 & 17.2 & 72.7\\
2014/11/19.54 & IAO105 (MITSuME) & R$_\mathrm{C}$ & 180 & 19  & 7.55 & 2.70 & 1.83 & 12.0 & 77.7\\
2015/01/29.58 & IAO105 (MITSuME) & R$_\mathrm{C}$ & 180 & 19  & 5.67 & 3.00 & 2.43 & 17.0 & 92.5\\
2015/01/30.46 & NHAO200 (MINT) & R$_\mathrm{C}$ & 180    & 7    & 5.64 & 3.01 & 2.45 & 17.1 & 92.7\\
2015/01/31.53 & OAO050 (MITSuME) & R$_\mathrm{C}$ & 120 & 43 & 5.60 & 3.01 & 2.46 & 17.2 & 92.9\\
2015/02/19.58 & IAO105 (MITSuME) & R$_\mathrm{C}$ & 180 & 28  & 4.96 & 3.10 & 2.78 & 18.4 & 96.4\\
2015/02/20.54 & IAO105 (MITSuME) & R$_\mathrm{C}$ & 180 & 47  & 4.93 & 3.10 & 2.78 & 18.4 & 96.5\\
2015/02/24.37 & BOAO180(4kCCD) & R$_\mathrm{C}$	& 180 & 5   & 4.83 & 3.11 & 2.86 & 18.4 & 97.2\\
2015/03/16.51 & OAO188 (KOOLS) & R$_\mathrm{C}$ & 180 & 36    & 4.28 & 3.20 & 3.22 & 17.8 & 100.7\\
\enddata
\tablecomments{The top header shows Exptime, the exposure time of each image in seconds; $N$, the number of images on a given date; $r_\mathrm{h}$, the heliocentric distance in AU; $\Delta$, the geocentric distance in AU; $\alpha$, the median phase angle; and $\theta_T$, the true anomaly in degrees. $\phi$ is an aperture diameter in arc-seconds to measure the area within the circle of radius 5,000 km from the center of the nucleus. We utilized the JPL Horizon Ephemeris program\footnote{http://ssd.jpl.nasa.gov/?horizons} based on the Web interface to obtain the median UT at the given observation date. The abbreviations of the telescope names denote the Siding Spring Observatory 50 cm telescope (SSO050), the Ishigakijima Astronomical Observatory 105 cm telescope (IAO105), the Okayama Astrophysical Observatory 188 cm (OAO188) and 50 cm telescopes (OAO050), the Nishi-Harima Astronomical Observatory 200 cm telescope (NHAO200), and the Bohyunsan Optical Astronomy Observatory 180 cm telescope (BOAO180).}

\end{deluxetable}

\clearpage
\begin{table}
\footnotesize
\caption{$Af\rho$ and Dust mass loss rate values of 17P/Holmes}
\begin{center}
\begin{tabular}{lrr}
\hline
UT date & $Af\rho$ [cm] & $\dot{M}_d$ [kg s$^{-1}$]\\
\hline
2013/05/04.67 & 67     & 0.61\\
2013/05/10.75 & 97     & 0.86\\
2013/05/20.54 & 103   & 0.87\\
2013/07/13.58 & 116   & 1.1\\
2013/07/31.58 & 121   & 1.3\\
2014/07/02.79 & 179   & 2.4\\
2014/07/04.79 & 177   & 2.4\\
2014/07/05.75 & 160   & 2.2\\
2014/07/17.75 & 126   & 1.8\\
2014/07/18.76 & 140   & 2.0\\
2014/07/24.81 & 185   & 2.6\\
2014/08/04.76 & 104   & 1.5\\
2014/08/06.79 & 99     & 1.4\\
2014/08/20.80 & 117   & 1.7\\
2014/08/22.79 & 95     & 1.4\\
2014/08/31.72 & 112   & 1.6\\
2014/09/03.71 & 89     & 1.3\\
2014/09/05.80 & 103   & 1.5\\
2014/09/10.73 & 96     & 1.4\\
2014/09/11.70 & 111    & 1.6\\
2014/09/13.67 & 102   & 1.4\\
2014/09/14.74 & 94     & 1.3\\
2014/09/15.70 & 92     & 1.3\\
2014/09/16.33 & 103   & 1.4\\
2014/09/20.75 & 88     & 1.2\\
2014/09/23.76 & 87     & 1.2\\
2014/09/26.75 & 87     & 1.2\\
2014/10/29.68 & 77     & 0.83\\
2014/11/19.54 & 56     & 0.50\\
2015/01/29.58 & 364   & 3.6\\
2015/01/30.46 & 344   & 3.4\\
2015/01/31.53 & 352   & 3.5\\
2015/02/19.58 & 74     & 0.75\\
2015/02/20.54 & 73     & 0.75\\
2015/02/24.37 & 55     & 0.56\\
2015/03/16.51 & 35     & 0.35\\ 
\hline
    \end{tabular}
  \end{center}
\label{tab:afrho}
\end{table}


\begin{thebibliography}{}
\bibitem[A'Hearn et al.(1984)]{A'Hearn et al.1984} A'Hearn, M.~F., Schleicher, D.~G., Millis, R.~L., Feldman, P.~D., \& Thompson, D.~T.\ 1984, \aj, 89, 579 
\bibitem[Altenhoff et al.(2009)]{Altenhoff et al.2009} Altenhoff, W.~J., Kreysa, E., Menten, K.~M., et al.\ 2009, \aap, 495, 975 
\bibitem[Boissier et al.(2012)]{Boissier et al.2012} Boissier, J., Bockel{\'e}e-Morvan, D., Biver, N., et al.\ 2012, \aap, 542, A73 
\bibitem[Drilling \& Landolt(2000)]{Drilling2000} Drilling, J.~S., \& Landolt, A.~U.\ 2000, Allen's Astrophysical Quantities, 381 
\bibitem[Holmes(1892)]{Holmes1892} Holmes, E.\ 1892, The Observatory, 15, 441 
\bibitem[Hillman \& Prialnik(2012)]{Hillman2012} Hillman, Y., \& Prialnik, D.\ 2012, \icarus, 221, 147
\bibitem[Hsieh et al.(2010)]{Hsieh et al.2010} Hsieh, H.~H., Fitzsimmons, A., Joshi, Y., Christian, D., \& Pollacco, D.~L.\ 2010, \mnras, 407, 1784 
\bibitem[Ishiguro(2008)]{Ishiguro2008} Ishiguro, M.\ 2008, \icarus, 193, 96 
\bibitem[Ishiguro et al.(2010)]{Ishiguro et al.2010} Ishiguro, M., Watanabe, J.-I., Sarugaku, Y., et al.\ 2010, \apj, 714, 1324 
\bibitem[Ishiguro et al.(2013)]{Ishiguro et al.2013} Ishiguro, M., Kim, Y., Kim, J., et al.\ 2013, \apj, 778, 19 
\bibitem[Jewitt \& Meech(1987)]{Jewitt1987} Jewitt, D.~C., \& Meech, K.~J.\ 1987, \apj, 317, 992 
\bibitem[Jewitt(2002)]{Jewitt2002} Jewitt, D.~C.\ 2002, \aj, 123, 1039 
\bibitem[Jewitt(2015)]{Jewitt2015} Jewitt, D.\ 2015, Nature Physics, 11, 96 
\bibitem[Kobayashi et al.(2013)]{Kobayashi et al.2013} Kobayashi, H., Kimura, H., \& Yamamoto, S. \ 2013, ana, 550, A72
\bibitem[Lamy et al.(2004)]{Lamy et al.2004} Lamy, P.~L., Toth, I., Fernandez, Y.~R., \& Weaver, H.~A.\ 2004, Comets II, 223 
\bibitem[Landolt(2009)]{Landolt2009} Landolt, A.~U.\ 2009, \aj, 137, 4186 
\bibitem[Levison \& Duncan(1997)]{Levison1997} Levison, H.~F., \& Duncan, M.~J.\ 1997, \icarus, 127, 13 
\bibitem[Li et al.(2011)]{Li et al.2011} Li, J., Jewitt, D., Clover, J.~M., \& Jackson, B.~V.\ 2011, \apj, 728, 31 
\bibitem[Luu \& Jewitt(1992)]{Luu1992} Luu, J.~X., \& Jewitt, D.~C.\ 1992, \aj, 104, 2243 
\bibitem[Meech \& Svoren(2004)]{Meech2004} Meech, K.~J., \& Svoren, J.\ 2004, Comets II, 317 
\bibitem[Miles(2010)]{Miles2010} Miles, R.\ 2010, arXiv:1006.4019 
\bibitem[Mink(1997)]{Mink1997} Mink, D.~J.\ 1997, Astronomical Data Analysis Software and Systems VI, 125, 249 
\bibitem[Reach et al.(2010)]{Reach2010} Reach, W.~T., Vaubaillon,  J., Lisse, C.~M., Holloway, M., \& Rho, J.\ 2010, \icarus, 208, 276 
\bibitem[Rickman et al.(1990)]{Rickman et al.1990} Rickman, H., Fern\'{a}ndez, J. A., \& Gustafson B. \AA. S. \ 1990, ana, 237, 524
\bibitem[Rosenberg \& Prialnik(2009)]{Rosenberg2009} Rosenberg, E.~D. \& Prialnik, D.\ 2009, \icarus, 201, 740 
\bibitem[Schulz et al.(2015)]{Schulz et al.2015} Schulz, R., Hilchenbach, M., Langevin, Y., et al.\ 2015, \nat, 518, 216 
\bibitem[Sekanina(2009a)]{Sekanina2009} Sekanina, Z. \ 2009, International Comet Quarterly, 31, 5
\bibitem[Snodgrass et al.(2006)]{Snodgrass et al.2006} Snodgrass, C., Lowry, S. C., \& Fitzsimmons, A. 2006, mnras, 373, 1590
\bibitem[Stevenson et al.(2010)]{Stevenson et al.2010} Stevenson, R., Kleyna, J., \& Jewitt, D.\ 2010, \aj, 139, 2230 
\bibitem[Stevenson \& Jewitt(2012)]{Stevenson et al.2012} Stevenson, R., \& Jewitt, D.\ 2012, \aj, 144, 138 
\bibitem[Stevenson et al.(2014)]{Stevenson et al.2014} Stevenson, R., Bauer, J.~M., Kramer, E.~A., et al.\ 2014, \apj, 787, 116 
\bibitem[Thomas et al.(2013)]{Thomas et al.2013} Thomas, P. et al. \ 2013, icarus, 222, 453
\bibitem[Whipple(1950)]{Whipple1950} Whipple, F.~L.\ 1950, \apj, 111, 375 
\bibitem[Whipple(1951)]{Whipple1951} Whipple, F.~L.\ 1951, \apj, 113, 464 
\bibitem[Yoshida(2005)]{Yoshida2005} Yoshida, M.\ 2005, Journal of Korean Astronomical Society, 38, 117 
\bibitem[Zacharias et al.(2010)]{Zacharias et al.2009} Zacharias, N., Finch, C., Girard, T., et al.\ 2010, \aj, 139, 2184
\end{thebibliography}
\end{document}